\documentclass[]{spie}  

\usepackage{aas_macros}
\usepackage{amsmath,amsfonts,amssymb,commath}
\usepackage{graphicx}
\graphicspath{{figures/}}

\usepackage[super]{nth}
\usepackage{siunitx}
\DeclareSIUnit{\parsec}{pc}
\DeclareSIUnit{\magnitude}{mag}
\DeclareSIUnit{\arcsec}{asec}
\DeclareSIUnit{\jansky}{Jy}
\DeclareSIUnit{\erg}{erg}
\DeclareSIUnit{\arcmin}{arcmin}
\DeclareSIUnit{\arcsec}{arcsec}
\DeclareSIUnit{\mas}{mas}
\usepackage[colorlinks=true, allcolors=blue]{hyperref}

\usepackage[capitalize]{cleveref}

\title{IQUEYE at Gemini South: instrument, science commission, and first results}

\author[a]{Tomas Cassanelli}
\author[a]{P. Marcone-Puga}
\author[b,c]{Giampiero Naletto}
\author[c]{Luca Zampieri}
\author[b, c]{Paolo Ochner}
\author[c]{Michele Fiori}
\author[c]{Alessia Spolon}
\author[d]{S. B. Araujo Furlan}
\author[e]{Albert Wait Kit Lau}
\author[f, g]{Ryan Mckinven}

\affil[a]{Department of Electrical Engineering, Universidad de Chile, Av.~Tupper 2007, Santiago 8370451, Chile}
\affil[b]{Department of Physics and Astronomy, University of Padova, Via F.~Marzolo 8, I-35131, Padova, Italy}
\affil[c]{INAF - Osservatorio Astronomico di Padova, Via Vicolo dell'Osservatorio 5, 35122, Padova, Italy}
\affil[d]{Instituto de Astronom\'{i}a Te\'{o}rica y Experimental, CONICET-UNC, Laprida 854, X5000BGR - C\'{o}rdoba, Argentina}
\affil[e]{Dunlap Institute for Astronomy and Astrophysics, University of Toronto, Toronto, Ontario, M5S 3H4, Canada}
\affil[f]{Department of Physics, McGill University, 3600 rue University, Montr\'eal, QC H3A 2T8, Canada}
\affil[g]{Trottier Space Institute, McGill University, 3550 rue University, Montr\'eal, QC H3A 2A7, Canada}

\begin{document} 
\maketitle

\begin{abstract}
  The Italian quantum eye (IQUEYE) is a fast photon counter based on the single photon avalanche diode detectors and capable of preserving a \qty{.5}{\nano\s\per\hour} accuracy photon time of arrival. IQUEYE was originally developed for intensity interferometry experiments, but now its scientific scope has been extended towards ultra fast astronomy, including optical pulsars, millisecond pulsars and the enigmatic fast radio bursts. IQUEYE's capabilities are mainly restricted by the number of photons detected, a quantity that scales with the collector size of an optical telescope. Through the visitor instrument program at Gemini South (Cerro Pachón, Chile) we brought IQUEYE to the 8.1-m dish, reaching an order magnitude sensitivity increased from previous operations. At Gemini South we installed IQUEYE to observe giant pulse emitters, millisecond pulsars, and transitional millisecond pulsars for over 40 hours in the span of a week. 
  Here we present the instrument and its adaptation to Gemini South interfaces, the instrument commission, and show the immediate first results from its early operations.
\end{abstract}


\section{INTRODUCTION}
\label{sec:introduction}  

Transient astronomical events are phenomena that appear and vanish within observable timescales ranging from milliseconds to days. These events can be broadly categorized into two groups: long-duration events, lasting from hours to decades, and short-duration events, occurring on seconds or sub-second timescales. In the optical domain, long-duration events, such as supernovae\cite{2025A&ARv..33....1R} and their evolving thermal emission, have been extensively studied. However, rapid short-duration events, particularly those lasting less than \qty{\sim1}{\milli\s}, remain largely unexplored.
The advancement of $\gamma$-ray detectors and the groundbreaking discovery of $\gamma$-ray bursts\cite{1998ApJ...497L..17S,1973ApJ...182L..85K}---still enigmatic in many aspects---have transformed the ability of optical telescopes to respond swiftly to transient phenomena, enabling rapid time-of-opportunity (RToO) observations and the tracking of potential simultaneous counterparts and/or afterglows. 

Since the early 2000s a family of new transient events have been discovered at long radio wavelengths, such as fast radio bursts\cite{2007Sci...318..777L} (FRBs), rotating radio transients\cite{2006Natur.439..817M} (RRATs), and magnetars' bright radio bursts \cite{2020Natur.587...59B,2020Natur.587...54C}. 
The latter, FRBs, are originated at extragalactic distances (cosmological applications\cite{2020Natur.581..391M}), and generally classified into repeating and one-offs sources (a single burst of millisecond duration). To date repeating sources account for less than \qty{10}{\percent} of all observed events with known periodicity\cite{2022ApJ...927...55L}, on- and off-activity cycles\cite{2020Natur.582..351C,2021MNRAS.500..448C} and other more complex scenarios.
Radio facilities have since achieved a sub-arcsecond localization of most repeating events\cite{2017Natur.541...58C,2022MNRAS.513..982R,2022Natur.606..873N,2024MNRAS.533.3174T,2024ApJ...977L...4H,2025ApJ...979L..21S}, leaving the vast majority of one-off events in a relatively large regions of the sky (generally a few arcminutes). Nevertheless, to enable multi-wavelength followups radio facilities are undergoing an expansion in localization capabilities\cite{2022AJ....163...65C,2024NatAs...8.1429C,2025arXiv250405192F} and soon hundreds of transients will be detected within \qty{100}{\mas}.

Transients have quickly populated the radio sky from $\nu=\qtyrange{.1}{6}{\giga\hertz}$ with over \num{1000} events discovered from Galactic to extragalactic distances\cite{2023Sci...382..294R} ($z=\numrange{0}{1}$). 
However, none of them have been detected at optical to near-infrared wavelengths\cite{2025MNRAS.538.1800H,2024ApJ...964..121K,2023HEAD...2011718H,2023A&A...676A..17T,2022ApJ...931..109N,2021A&A...653A.119N,2018ApJ...860...73E,2017MNRAS.472.2800H} (ONIR), despite the multiple theories proposing their existence\cite{Yang_2019,2014PASJ...66L...9N}.
The absence of detections can be attributed to the challenge of monitoring a highly specific and narrow field-of-view (FoV; just a few arcseconds) at precisely the right moment, and the instrument limitation to fully resolve such rapid events (smeared out). 
%
Besides instrument sensitivity, this limitation is driven by the number $N_\text{p}$ of photons collected\cite{1999ASPC..180..671R} at ONIR wavelengths. 
%
Therefore, the quest for ultra rapid transient events must, at an initial stage, be pursued on localized and known to repeat sources, coupled with an ultra fast detector and large collector.


The \textit{Italian quantum eye}\cite{2009A&A...508..531N} (IQUEYE)\footnote{\url{https://web.oapd.inaf.it/zampieri/aqueye-iqueye}.} is one of the leading instruments at very rapid timescales in the optical band. The system works with four single photo avalanche diodes (SPADs) detectors and a time-tagging acquisition system that includes a Rubidium oscillator (clock) and a GPS (global positioning system) based receiver (UTC referenced), i.e., it can time-tag single photons with a time accuracy better than \qty{.5}{\nano\s\per\hour} (\qty{.5}{\nano\s} accuracy over \qty{1}{\hour} of free running clock). 
The instrument has no spatial imaging capability and a small FoV \qtyrange{1}{2}{\arcsec} (depending on the pinhole aperture used), hence targets require arcsecond localization. IQUEYE has been previously used in optical facilities, but it has never been placed in a 8-m class telescope such as Gemini South\cite{2010SPIE.7735E..05T} until now. In addition, no current instruments are capable of achieving IQUEYE's timing accuracy, and limited to within \qty{\sim10}{\milli\s} exposures\cite{2021FrASS...8..138S}.
From IQUEYE earlier results, multiple studies have been performed observing pulsars and transients\cite{2022A&A...663A.106C,2019MNRAS.482..175S,2014MNRAS.439.2813Z} and studying the coherence of star light in intensity interferometry experiments\cite{2021MNRAS.506.1585Z,2021SPIE11835E..0DF}. Now, poised to become a lead instrument in the rapidly changing field of ultra fast astronomy (UFA), IQUEYE will probe pulsars and millisecond pulsars to validate its scientific capabilities, and in the near future perform dedicated searches for other puzzling transient phenomena. IQUEYE, a visiting instrument at Gemini South, will pave the way for future UFA science and help in the development of future instrumentation with this scientific goal in mind.

This paper describes the instrument commission and first science results of IQUEYE at Gemini South. \Cref{sec:interfaces} describes Gemini South interfaces (mechanical, optical, power, and software) as well as how it was adapted to the telescope, \cref{sec:observing_program} describes our observing program and first sources targeted by IQUEYE, and \cref{sec:early_science_results} shows our first science results. Supplemental material in the expected flux and counts of our experiment in shown in \cref{sec:system_response}. Lastly in \cref{sec:conclusions} conclusions and prospects are presented.

\section{IQUEYE DESIGN AND ADAPTATION TO GEMINI SOUTH}
\label{sec:interfaces}

IQUEYE along its sibling instrument AQUEYE+ (\textit{Asiago quantum eye}) have been performing intensity interferometry experiments since late-2000s, and have only been exposed to European traditional optic configurations, e.g., Telescopio Nazionale Galileo (\qty{3.58}{\m}; Spain), New Technology Telescope (NTT \qty{3.58}{\m}; Chile), and Galileo telescope (\qty{122}{\centi\m}; Italy).
IQUEYE was originally developed to fullfil an ($f$-number) $f/2.2$ optics standard, later adapted to work using an optical fiber link, but not made to operate at a $f/16$ focal ratio (Gemini South configuration). Nevertheless, early simulations estimated few losses when coupling IQUEYE with an optimized opto-mechanical flange (\cref{sec:optomechanical}), and an overall improved sensitivity can be achieved, regardless the original configuration. 
Instrument electronics (\cref{sec:eletronics}) required preparation for a completely remote access, and software (\cref{sec:software}) minimal modifications. This section describes the IQUEYE's adaptation to Gemini South.

\subsection{Opto-mechanical adaptation}
\label{sec:optomechanical}

\begin{figure}[t]
  \centering
    





  \includegraphics{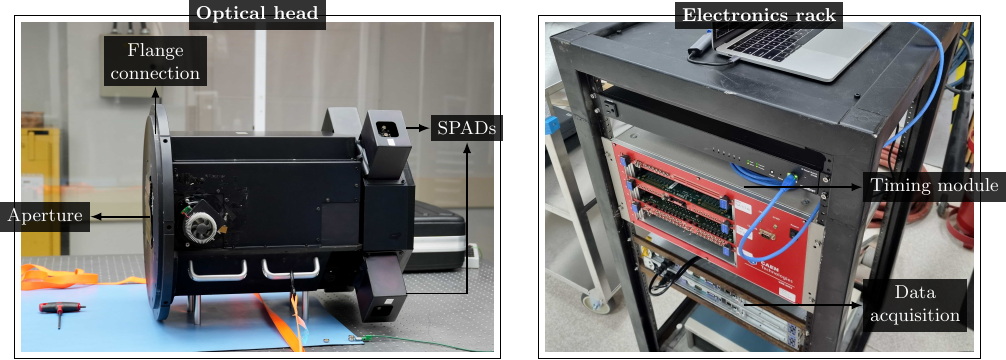}
  \caption{\textit{Left}: IQUEYE optical head at Gemini South. The optical head is composed of a selectable pinhole, two filter wheels (including polarizers), demagnification optical components, and a pyramid to split light into the four SPADs. IQUEYE's instrument complete description in Naletto et al\cite{2009A&A...508..531N}. SPADs are located at the back of the optical head and they are directly connected to a power source and a data transfer cable. These connections run towards the electronics rack. \textit{Right}: IQUEYE's electronics rack. Photograph shows the rack with no cable connections. The rack hosts the power distribution unit, timing modules, data acquisition, and Rb clock and GPS modules. All equipment is hosted and then remotely accessed through a single computer and within Gemini South's control room. Both photographs show instrument being commission on site, February 2025.}
  \label{fig:iqueye}
\end{figure}

The optical head (\cref{fig:iqueye}) guides the light through its aperture, filters, demagnification (of a factor \num{3.25}), and then towards a pyramid which splits the light in four paths. SPADs are located at the end of the optical train with an effective detecting area of \qty{7800}{\micro\m\squared} (diameter $d=\qty{100}{\micro\m}$). SPADs detectors are commercial components with an advertized photo-detection efficiency\cite{2016JInst..11P8014Z} (PDE) of \qty{50}{\percent} at $\lambda=\qty{500}{\nano\m}$. PDE is usually expressed in terms of:
\begin{equation}
  \text{PDE} = \text{FF} \times \text{QE} \times T_\text{p}. \label{eq:pde}
\end{equation}
Here, FF denotes the fill factor (the ratio between the total detector area and the active area), QE is the quantum efficiency (the probability that an incident photon generates an electron-hole pair in silicon that can be detected), and $T_\text{p}$ represents the trigger probability, i.e., the likelihood that a primary carrier initiates an avalanche in the cell.

Because of the detector size, two stages of demagnification were included in the optics to accommodate an optimum plate-scale matching. However, this optimization was done for a different $f$-number, and to understand consequences we simulated losses on ANSYS OpticStudio, (IQUEYE coupled with Gemini South; \cref{sec:simulations}).

\subsubsection{Simulations}
\label{sec:simulations}

\begin{figure*}[t]
  \centering
  \includegraphics{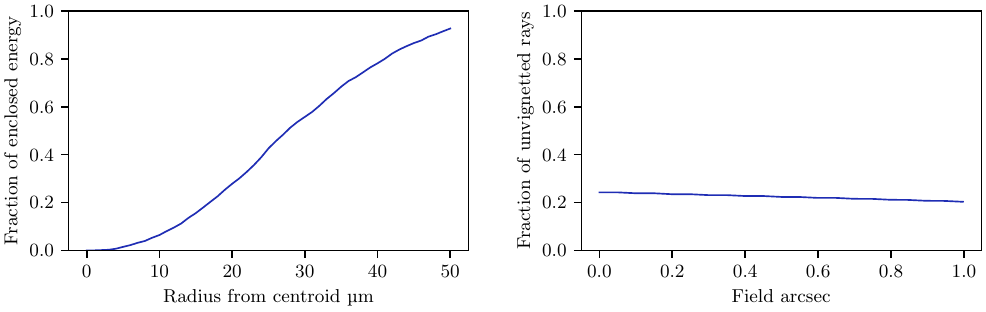}
  \caption{ZEMAX optics simulations. \textit{Left}: enclosed energy or the efficiency of the system coupling IQUEYE with Gemini optics. The maximum is reached at \qty{92.7}{\percent}. \textit{Right}: fraction of unvignetted rays against the observed field for IQUEYE at Gemini South simulation. Notice that this result corresponds only to a single SPAD and the total fraction needs to be multiplied by a \num{4} factor.}
  \label{fig:zemax_results}
\end{figure*}

We simulated standard seeing conditions with a \qty{2}{\arcsec} diameter FoV input to Gemini South, with an extended circular uniform source of \qty{1}{\arcsec} radius at the center of the FoV and four point sources at the extremes of the aperture. Due to the $f$-number difference, we see that lenses within IQUEYE are partially filled and light is in a decentered position. However this is not a substantial impact on the final spot quality. Spot diagrams are practically within the SPAD diameter (\qty{100}{\micro\m}), with some losses only at the longest wavelengths because of chromatic aberration residuals. Quantitatively we computed the fraction of enclosed energy with this configuration, summarized in \cref{fig:zemax_results}, where we see that the system achieves over a \qty{\sim90}{\percent} of the energy within the SPAD diameter.
In practice, the only minor critical optics component is the pyramid tip which is flattened over a square of \qty{1}{\milli\m} (pyramid and optical components described elsewhere\cite{2009A&A...508..531N}). This is not considered in \cref{fig:zemax_results} left plot. To properly address this effect we can look at the vignetting effect shown in \cref{fig:zemax_results} right plot. The maximum and minimum values are \qty{24.2}{\percent} and \qty{20.3}{\percent}, which has only been computed for a single SPAD arm (i.e., the total throughput must be multiplied by a factor of \num{4}), viz., the total unvignetted rays are about \qty{88}{\percent}.

Lastly we evaluated the effect on slew/pointing stability of Gemini South on our observations, assuming a value of \qty{15}{\mas\per\s}. We run again our simulations with this same \qty{1}{\arcsec} radius extended uniform source and the fraction of enclosed energy lost is less than \qty{0.4}{\percent} (in \cref{fig:zemax_results} left from \qty{92.7}{\percent} to \qty{92.3}{\percent}).

With this simulation we can conclude that the $f$-number difference was not a problem, and no further optical modifications were performed to the system. As mentioned, the optical performance will not be optimal, but it is sufficient for Gemini's characteristics and scientific purpose of the instrument. 

IQUEYE has access on its optics to three (selectable) pin-hole sizes, which can be adjusted to reduce background of observations, these are: \qtylist{200;300;500}{\micro\m}. In addition, the internal optics of the instrument have a magnification factor of \num{3.25}. Then the (estimated) combined optics of Gemini South and IQUEYE (current configuration) return FoVs are of the order of \qtylist{1.04;1.56;2.6}{\arcsec}. For our observations we opted for the largest setting to have a better pointing calibration of the commissioned instrument.

\subsubsection{Mechanical adaptation}

\begin{figure*}
  \centering





  \includegraphics{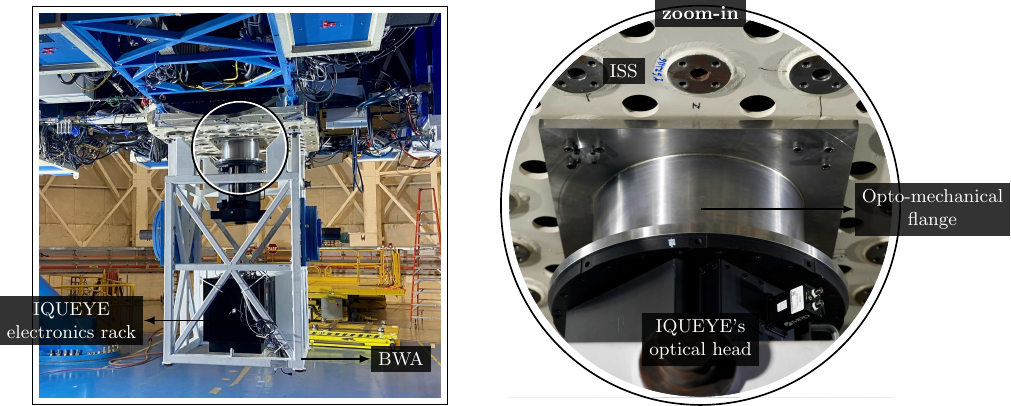}
  \caption{IQUEYE at Gemini South. From top to bottom (left-hand side): ISS, flange, and electronics rack. The white mechanical structure is the ballast weight assembly (BWA), a structure to give balance to the telescope and attach any remaining instrument load (such as the electronics). The total weight of the system BWA, flange, IQUEYE optical head and electronics is \qty{2005.5}{\kg} and the entire system has a precisely tunned center of mass near the center of the BWA. Posterior the install of the whole structure a mechanical balance must be done to the entire telescope.
  The optical head is independently attached to the ISS from the BWA, thus, mechanical stress and perturbations to optical light path come only from the flange and optical head contributions. In the {\textbf{zoom-in}} section (right-hand side) the top section of the optical head and flange are shown attached to the ISS. Photograph taken on February 2025.}
  \label{fig:iqueyeatgemini}
\end{figure*}

Because of the instrument's optical head simplicity, size (\qtyproduct{\sim60x60x50}{\centi\m}), and weight \qty{130}{\kg}, its adaptation to Gemini South required solely a flange. The flange is then directly connected to Gemini's Cassegrain optical focus, and opto-mechanically attached to the instrument support structure (ISS). The ISS is a stiff cube that hosts up to five instruments, being the bottom section the largest to accommodate instruments. The Cassegrain focus is located at a distance of \qty{300}{\milli\m} from the ISS surface. With this specification we then designed the flange and centered IQUEYE's optical head. An independent additional mechanical structure is projected from one of the ISS faces, the ballast weight assembly (BWA). The BWA function is to host any remaining instrument component (e.g., electronics) and to add balance to the whole telescope structure. Nevertheless, each section BWA and flange are independently connected to the ISS and they do not influence each other (no mechanical stressed shared), therefore optics remain unaffected.
Each instrument has its own ballast weight assembly (BWA) mechanical structure and designed to specifications. \Cref{fig:iqueyeatgemini} shows the arrangement of each of these opto-mechanical components. 

The flange, made of solid aluminum A36 (\qty{89.4}{\kg}), attaches directly to the ISS and to IQUEYE's optical head keeping all movement and shear stress isolated from additional components. The flange was designed over an iterative process, including simulation results and a mechanical analysis of the system. The final design was then manufactured and assembled at the Gemini South facility one day before the observations.

\subsection{Electronics adaptation}
\label{sec:eletronics}

IQUEYE electronics start from the analog signal coming from the SPADs (attached to IQUEYE's optical head) and then feed to a field-programmable gate array (FPGA). At a later stage, a time-tagging subsystem tags in all bit-converted photons from the FPGA and stored to disk. Data rates are minimal and there is no risk of overflow\cite{2009A&A...508..531N}. After observations data are copied and transferred over Ethernet for post-processing. The time-tagging needs for IQUEYE further exceed the current timing applications at Gemini South (to within \qty{\sim10}{\milli\s}), and an independent timing system was included. The time of references is achieved throughout two independent systems: a Rubidium ovenized clock and a GPS receiver (composed of a receiver and antenna). 

All IQUEYE electronics, including data acquisition system (two rack size computers), FPGAs, Rubidium clock, and GPS receiver; are housed within the electronics rack (\cref{fig:iqueyeatgemini}). The system, not designed for a fully remote access, was upgraded for a full distant action with a power distribution unit and an uninterrupted power supply. Then all electronics can be accessed through Ethernet from Gemini South control room (city of La Serena) and no additional modifications were needed.

Only one component must be outside the electronics rack, the GPS antenna. For IQUEYE's commission the GPS antenna was installed within the telescope dome, near the main collector and in an azimuth dependent location. The antenna position directly affects the timing residuals, nevertheless, a time correction will be applied in a post-processing stage. For future observations a fixed location will be chosen, far away from the dome (where amplification of the analog signal will be required). At all times, the GPS antenna had on sight three or more satellite calibrators, enough to provide a stable timing solution for our commission observations.

The electronics rack does not include any cooling since all electronics did not reach the maximum allowed by Gemini South. Lastly all commercial lights from electronics components were removed or covered to not alter telescope operations.

\subsection{Software adaptation}
\label{sec:software}

IQUEYE, a single pixel instrument, has a single parameter to monitor its behavior, the number of photon counts (source flux). Based on this we know whether the instrument is pointing in a dark count setting, with filters, or on- and off-source. Since this information is rather manipulated by the observer no additional software interfaces needed to be developed to include it in the Gemini South system.
No further adaptions were required at this stage. Future implementation will retrieve some of the telescope information to fill in data headers, nevertheless a realtime interaction of instrument and telescope is no necessary for IQUEYE's operations.


\section{Observing program}
\label{sec:observing_program}

As discussed in \cref{sec:interfaces}, all instrumentation was accessed remotely with a few components that remained at all times connected to power. Nights started with darks calibration, flux calibration, and then for each source an instrument offset pointing calibration followed by the scientific source of interest (target).

Dark counts are performed at the beginning of the night in addition to a few flux calibration sources (bright stars) at twilight and dawn of the observations. At all times we used the largest pin-hole configuration, i.e., a FoV of \qty{2.6}{\arcsec}.
Telescope pointing and adjustments were performed independently of IQUEYE, and following the standard routines of Gemini South.

Source location was first done with the acquisition and guidance\cite{1998SPIE.3355..206H} (A\&G) system within Gemini South (which blocks light to IQUEYE while on use) and its pointing software. IQUEYE was then used only to confirm the number of observed counts of a pointing calibrator. Once on-calibrator source (a bright star within the A\&G FoV) an iterative process of on- and off-source is performed based on the number of counts obtained. Then an instrument offset is computed, given by the total counts we apply. Lastly, a blind offset is triangulated and applied to the centered calibrator towards the target (similarly as done with the GHOST\cite{2024AJ....168..208K} instrument). The process was then repeated for every field within a target. No additional pointing calibration was performed to the system (e.g., no spiral calibration search).

Our observing program consisted of ordinary optical pulsars (Crab, Geminga, Crab twin, and Vela\cite{2000Msngr..99...22M}), millisecond pulsars , transitional millisecond pulsars (PSR J1227$-$4853 and PSR J1023$+$0038), and known giant pulse emitters at X- and $\gamma$-rays (PSR J1231$+$1411 and PSR J1823$-$3021A). An observation summary is listed in \cref{tab:targets}. In addition, multi-wavelength observations were performed simultaneous and/or contemporaneous to these set of targets, including the Algonquin Radio Observatory (ARO; 46-m dish $\nu=\qtyrange{400}{800}{\mega\hertz}$), the Instituto Argentino de Radioastronomía (IAR) 30-m dish ($\nu=\qtyrange{1200}{1600}{\mega\hertz}$), the Green Bank Telescope (GBT; \qtyrange{1.15}{1.73}{\giga\hertz}), and Neutron Star Interior Composition Explorer Mission (NICER; \qtyrange{0.2}{12}{\kilo\electronvolt}).

\begin{table*}
  \centering
  \caption{Sources observed with IQUEYE at Gemini South during February 15--18, 2025. The program observed effectively \num{26.6} hours (reduction of the \qty{50}{\hour} include commissioning and pointing/calibration to targets). The visual magnitude $m_V$ is listed in the second column using reported values in the literature.}
  \label{tab:targets}
  \begin{tabular}{l|l|llll|l}
    \hline
    PSR Source & $m_V$ & 15/02/2025 & 16/02/2025 & 17/02/2025 & 18/02/2025 & Total \\ \hline
    J0534$+$2200 (Crab) & \num{16.6 +- .1}\cite{1969Natur.221..525C,1996A&A...314..849N} & \qty{0.75}{\hour} & \qty{1.5}{\hour} & \qty{1}{\hour} & \qty{.5}{\hour} & \qty{3.75}{\hour} \\
    J0633$+$1746 (Geminga) & \num{26 +- .4}\cite{1998A&A...335L..21S} & \qty{1.5}{\hour} & \qty{1.33}{\hour} & -- & \qty{2.7}{\hour} & \qty{5}{\hour} \\
    J0540$-$6919 (Crab twin) & \num{22.26(0.2)}\cite{1985Natur.313..659M} & \qty{1}{\hour} & -- & \qty{1}{\hour} & \qty{2.5}{\hour} & \qty{4.5}{\hour} \\
    J0835$-$4510 (Vela) & \num{23.7 +- .5}\cite{1976ApJ...203..193L} & \qty{1}{\hour} & \qty{1}{\hour} & \qty{1.33}{\hour} & \qty{.75}{\hour} & \qty{4}{\hour} \\
    J1227$-$4853  & -- & -- & \qty{2}{\hour} & \qty{1.5}{\hour} & -- & \qty{3.5}{\hour} \\
    J1231$-$1411  & -- & -- & -- & \qty{1.33}{\hour} & -- & \qty{1.33}{\hour} \\
    J1023$+$0038  & $\num{17.3}^{+0.5}_{-0.8}$\cite{2020MNRAS.498L..98B,2017NatAs...1..854A} & \qty{.5}{\hour} & \qty{1}{\hour} & -- & \qty{.5}{\hour} & \qty{2}{\hour} \\
    J1823$-$3021A & -- & \qty{1}{\hour} & -- & -- & \qty{1.5}{\hour} & \qty{2.5}{\hour} \\ \hline
  \end{tabular}
\end{table*}

\section{Early science results}
\label{sec:early_science_results}

\begin{figure}[t]
  \centering
  \includegraphics{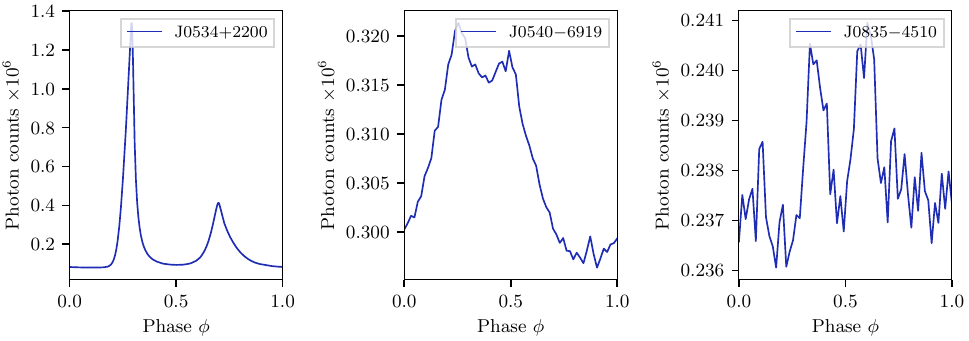}
  \caption{Pulsars' folded profile for PSR J0534$+$2200, PSR J0540$-$6919, and PSR J0835$-$4510. Folded profiles from left to right were produced with samples per period of: \num{512} ($\Delta{t}=\qty{66}{\micro\s}$), \num{64} ($\Delta{t}=\qty{795}{\micro\s}$), \num{64} ($\Delta{t}=\qty{1.4}{\milli\s}$). Observations correspond to a transit of only \qty{30}{\minute} for each source showing the number of counts of the folded profile. Each pulse shape was processed and adapted to the radio standard software PRESTO\cite{2011ascl.soft07017R}, and only including their first order parameters, such as period ($1/f$) and its derivative.} 
  \label{fig:folded}
\end{figure}

Our first observations targeted pulsars, known to emit at optical wavelengths, such as the Crab and Vela. The latter was observed at the beginning of every night (\cref{tab:targets}) and was used to confirm and validate our system sensitivity prior targeting weaker and/or never observed in the optical band sources. Because of the nature of photon counters, an initial increment in photon flux is obtained (on- and off-source) but only after data pre-processing we can visually confirm the source detection, either through single pulses or folded profiles (otherwise nebula emission or simply the noise background could be misinterpreted).
\Cref{fig:folded} shows the folded profiles of three sources, PSR J0534$+$2200 (Crab), PSR J0540$-$6919 (Crab twin), and PSR J0835$-$4510 (Vela). For the three sources a strong pulse shape could be seen with a few minutes of integration (particularly for the Crab pulsar), where a precise Earth location coordinates, up-to-date ephemerides, and pulsar properties (e.g., period $1/f$; $\phi\del{t} = \phi_0 + f\del{t-t_0} + \dot{f}\del{t-t_0}^2/2 + \dotso$) were included to optimize their optical folded pulse profile. These coupled with IQUEYE's superb timing accuracy retrieved the {highest signal-to-noise ratio (SNR) of folded profiles ever recorded at optical wavelengths given a time window of only \qty{30}{\minute} (for each pulsar shown in \cref{fig:folded}).} 
The technique of high cadence observations coupled to folding the observing time over the pulsar period allows to improve the pulsar SNR: in fact, from one side the signal (pulsating component) increases linearly because it is summed in phase, on the other the noise, that is the background, increases with the square root because of its statistically Possonian dependence. In addition, in some special cases as with the Crab pulsar which has narrow peaks and negligible background, this provides a signal whose magnitude, over the peak time intervals, is larger than what is reported in literature (\cref{tab:targets}) for that object. In fact, all the pulsating component is now concentrated on the peaks, and not smoothed over the whole pulsar period.
%
%
In general, these high cadence optical observations are limited by the available number of photons that we can collect. A star at zero magnitude $m_V=0$ will emit \qty{\sim1007}{\per\centi\m\squared\per\s\per\micro\m} photons per area, per time, and per passband $\Delta\lambda$\cite{2009ASSL..360.....C}, therefore at Gemini South, thanks to its 8.1 m diameter aperture, we expect 1 photon every \qty{2}{\nano\s} (considerably less for a greater magnitude). This means that, statistically, no losses in photons come from the instrument cadence and readiness to receive the next incoming photon, \qty{.5}{\nano\s\per\hour} (system response described in \cref{sec:system_response}).

\begin{figure}[t]
  \centering
  \includegraphics{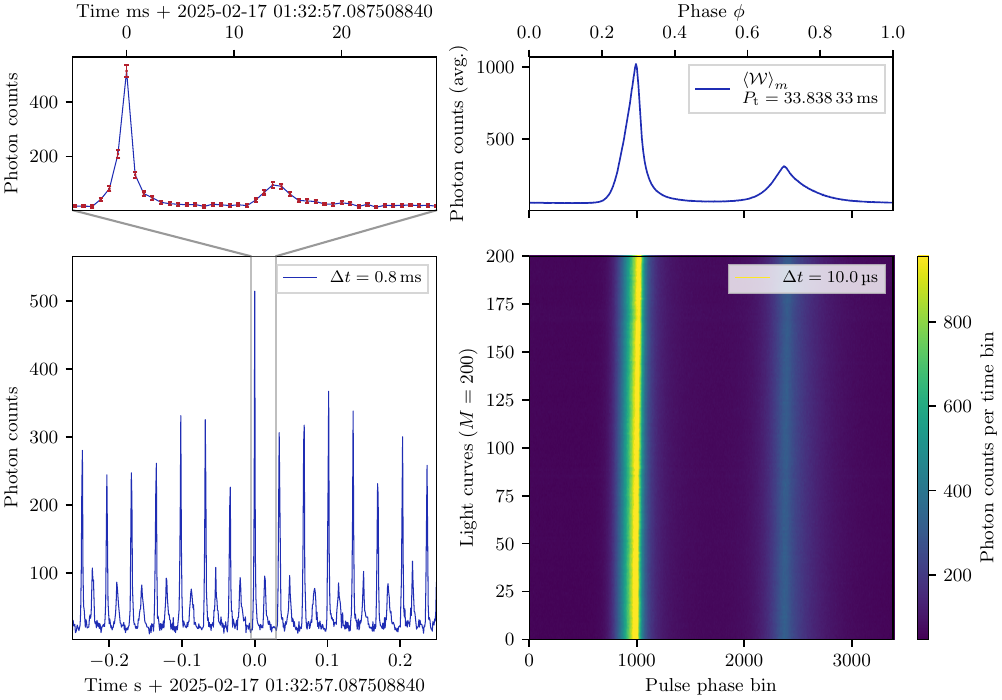}
  \caption{Crab pulsar (PSR J0534$+$2200) observed at Gemini South with IQUEYE. \textit{Left}: Train of pulsar pulses of the Crab pulsar (same data as in the right) of \qty{0.5}{\s} long. Data has a time bin of $\Delta{t}=\qty{0.8}{\milli\s}$, where a clear pulse profile of the main and interpulse are most of the time visible and above the noise floor. The top section corresponds to the zoom-in (gray lines in bottom plot) of a full pulsar rotation (this is \qty{33.838}{\milli\s}), where the pulse structure can be clearly identified (error bars are simply $\sqrt{\text{counts}}$).
  \textit{Right}: Waterfall diagram $\mathcal{W} = \del{w_{m,n}}\in\mathbb{R}^{M\times N}$ of \qty{30}{\minute} of data. The diagram is built by folding using the trial period $P_\text{t}=\qty{33.83822}{\milli\s}$, $N=\text{round}\sbr{P_\text{t}/\Delta{t}}=\num{3384}$, over $M=\num{200}$ continuos sections of the data stream (number of divisions). The top panel shows the average along $m$-axis (summing over $M$ rows), retrieving the standard pulsar folded profile at optical wavelengths. The waterfall plot has been generated using a fixed trial period $P_\text{t}$ therefore a perfect straight line may not be possible since higher order effects have not been considered. Similar figures have been shown in the past at smaller collector size\cite{2022A&A...663A.106C} but the current presents an overall improvement in signal strength.
  }
  \label{fig:crab_iqueye}
\end{figure}

Simultaneous radio observations were performed with Crab and Vela pulsars and analyzes here will compare their time of arrival or even for the very first time compare pulse structure at optical wavelengths. This last point can be seen in \cref{fig:crab_iqueye} were the full train of pulsar pulsations can be observed (after applying a time bin of $\Delta t=\qty{.8}{\milli\s}$). Such observations can push the limit to within $\Delta{t}\approx\qty{20}{\micro\s}$.
\Cref{fig:crab_iqueye} also shows the improved SNR of the Crab compared to previous measurements performed by IQUEYE\cite{2009A&A...508..531N,2022A&A...663A.106C} (or even its sibling instrument AQUEYE+\cite{2012A&A...548A..47G}), approaching the expected \num{5}-times factor in photon counts given by the collector size (from a \qty{3.58}{\m} to \qty{8.1}{\m} telescope). \Cref{fig:crab_iqueye} also shows for the very first time the pulse structure of an isolated pulse from the Crab pulsar, hence its highest flux detection to date at optical wavelengths.

Multiple studies convening these observations are underway, particularly when comparing pulse to pulse, average pulse width across observations, energy distribution of pulses (similarly to what has been done at other wavelengths\cite{2019MNRAS.490L..12B}), potential evidence of modulation, etc. Further, a multiwavelength analysis may open the very first comparison of radio-to-optical pulse identification, and see whether such properties are correlated within the pulsar itself or its magnetosphere.

Observations in \cref{fig:folded,fig:crab_iqueye} are in a preliminary state, were an additional corrections may be required. Flux calibration and other astronomical calibration properties remain to be included.

\section{Total system response of IQUEYE at Gemini South}
\label{sec:system_response}

To cross check our results with IQUEYE's we performed a first order estimation of the combined IQUEYE and Gemini South expected photon counts, in terms of the total system response.

The number of photons per second\cite{2005PASP..117..421T} collected is given by:
\begin{align}
  N_\text{p} = \int F_\lambda\del{\lambda^\prime} S\del{\lambda^\prime} \frac{\dif\lambda^\prime}{h\nu} = \frac{1}{hc} \int \lambda^\prime F_\lambda\del{\lambda^\prime} S\del{\lambda^\prime} \dif{\lambda^\prime},
\end{align}
where $\lambda$ is the observing wavelength within a passband $\Delta\lambda$, $F_\lambda$ is the intrinsic spectral energy distribution (equivalent to the specific flux density at radio wavelengths, and typically measured in \unit{\watt\per\m\squared\per\micro\m}), and $S$ is the total system response:
\begin{align}
  S\del{\lambda} = T\del{\lambda, z}Q\del{\lambda}H\del{\lambda}A.
\end{align}
Classically $N_\text{p}$ is different from zero only during the opening and closing of the shutter which lasts $\tau$ (viz., the total number of photons is given by $n_\text{p}= \tau N_\text{p}$).
The total system response $S$ will depend on multiple distributions, the atmospheric transmission $T$ (which will also be weighted over the source location and sky brightness; $z$ zenith angle), the telescope and instrument efficiencies (light throughput, e.g., reflection losses, PDE) $Q$, the filter response function $H$ and the collector area $A$.

Here we derived a first order approximation of the expected photon rate, $n_p$, on ideal conditions for IQUEYE at Gemini South. Our wavelength coverage (instrument passband $\Delta\lambda$) will be given as a tophat distribution limited at $\Delta\lambda$, and we approximate $S$ as a constant value for our centered wavelength $\lambda_c = \del{\lambda_f - \lambda_i}/2$,
\begin{align}
  n_p &\approx \frac{1}{hc} {T\del{\lambda_c, z} Q\del{\lambda_c} R\del{\lambda_c} A} \int_{\lambda_i}^{\lambda_f} {\lambda^\prime}^2\dif{\lambda^\prime} \approx \frac{1}{hc} {T\del{\lambda_c, \theta} Q\del{\lambda_c} R\del{\lambda_c} A} \frac{1}{2}\del{\lambda_f^2 - \lambda_i^2}, \\
  n_p &\approx \frac{1}{2hc}  \sbr{X\del{z}\times K\del{\lambda_c}} \times \sbr{\text{PDE}\del{\lambda_c} \times R\del{\lambda_c} \times (1 - p_\text{fp})} \times A\del{\lambda_f^2 - \lambda_i^2}. \label{eq:Np_number}
\end{align}
Where we have expressed $T\del{\lambda_c, \theta} = X\del{z}\times K\del{\lambda_c}$ (airmass and extinction), $Q = \text{PDE}\del{\lambda_c} \times R\del{\lambda_c}$ (PDE, reflection efficiency, and false positive probability), and $H\del{\lambda_c} = 1$ (since no filters were used for most observations). 
Afterpulsing effects and dark counts probability\cite{2013JInst...8P5010G,10.1117/1.JATIS.11.2.028007} are considered simply as $p_\text{fp} = \qtyrange{.1}{3}{\percent}$, based on SPADs manufacture. That is the probability of detecting an extra noise count, originated by residual charge or re-emission during avalanche amplification, hence a surplus would mean $(1 + p_\text{fp})$.

\begin{table}[t]
  \centering
  \caption{Combined Gemini South and Iqueye properties.}
  \label{tab:gsiqueye}
  \begin{tabular}{lll}
    \hline
    \textbf{Variable} & \textbf{Value} & \textbf{Description} \\ \hline
    Passband & $\Delta\lambda = \qtyrange{420}{720}{\nano\m}$ & Effective observing passband \\ 
    Extinction at $\lambda_c=\qty{550}{\nano\m}$ & \qty{.89}{\percent} (\num{.12} mag / airmass) & Median extinction coefficient\cite{1998SPIE.3355..614B} \\ 
    False positive probability & $p_\text{fp}=\qtyrange{.1}{3}{\percent}$ & Afterpulse and dark counts \\
    \hline
  \end{tabular}
\end{table}

%
We made these calculations using several known flux calibrator stars, and all the observations from PSR J0534$+$2200 in a on- and off-source strategy to subtract the sky background counts from each source.
We computed then all components in \cref{eq:Np_number} using standard values listed in \cref{tab:gsiqueye}, leaving free PDE and $R$ (plus other losses hard to derive or coupled) and compared the expected theoretical $n_p$ (at a theoretical \qty{100}{\percent} with a time bine of $\Delta{t}=\qty{1}{\s}$) and the observed IQUEYE counts.

We made these calculations using several known magnitude calibrator stars in an on- and off-source strategy to subtract the sky counts from sources. We computed then all components in \cref{eq:Np_number} using standard values listed in \cref{tab:gsiqueye}, leaving free PDE and $R$ (plus other losses hard to account) and measure the best fitting value for $R$. We then compared the expected theoretical $n_p$ (at a theoretical \qty{100}{\percent} with a time bin of $\Delta{t}=\qty{1}{\s}$) for every PSR J0534$+$2200 acquisition against the observed IQUEYE counts.
Assuming the worst case of \qty{3}{\percent} afterpulse probability, we ran this calculation for \num{6} calibrator stars with a total of \num{13} different observation acquisitions to compute. This resulted in a measured expected mean $R$ value of \qty{25 \pm 9}{\percent}.  Then, using the \num{13} individual acquisitions made in this run for PSR J0534$+$2200. We can compare the counts expected using this $R$ value, which resulted in a prediction error of \qty{0.249 \pm 0.1}{\magnitude}. Meaning we can predict counts for a source within that margin.
In a future implementation, we expect to run several magnitude and PDE calibrations to correctly model our system, this way, the $R$ value would be measured precisely, and therefore, the counts predictions can be improved with a lower error margin.

\section{Conclusions \& Future developments}
\label{sec:conclusions}

We have installed, commissioned, and scientifically operated the very first fast photon counter for UFA at Gemini South. IQUEYE is an instrument built for other applications in mind, and not optimized for Gemini South optics. Here we showed that its applicability is feasible, obtaining remarkable and promising results during its first observing campaign. These results include the highest SNR from folded profiles of multiple pulsars at optical wavelengths (given a fixed time window), and the very first fully resolved optical pulse profile of a pulsar (\cref{fig:crab_iqueye}). Our future science will continue surveying this new vastly unexplored parameter space for pulsars and include southern repeating (and active) FRB sources or other Galactic transient phenomena such as magnetars.
Lastly, IQUEYE will remain for future semesters at Gemini South as a resident visiting instrument, where our team will support the astronomical community observations.

\acknowledgments 

T.C.\ gratefully acknowledges support by the ANID BASAL FB210003. T.C.\ IQUEYE and team gratefully acknowledges the support of Gemini South staff, mechanical team, electronics, software and operations for their invaluable support, expertize and help before and during instrument commissioning.
The IQUEYE@Gemini project has been funded by the Department of Physics and Astronomy of the University of Padova, with a grant of the PARD 2024 Call.
GN, PO and LZ wish to thanks all the personnel of the Asiago observatory who contributed to the IQUEYE@Gemini project.
We acknowledge financial support from the INAF Research Grant ``Uncovering the optical beat of the fastest magnetized neutron stars (FANS)'' (PI: A. Papitto) and ``Coordinated Multiwavelength Exploration of Fast Radio Bursts (COMEFAR)'' (PI: M. Pilia).
The Dunlap Institute is funded through an endowment established by the David Dunlap family and the University of Toronto.
S.B.A.F. is a PhD candidate with a CONICET fellowship.
 
\bibliography{gs-iqueye} 
\bibliographystyle{spiebib} 

\end{document}